\documentclass{PoS}
\usepackage{amsmath}

\title{Low-energy constants from ALEPH hadronic tau decay data}

\ShortTitle{Low-energy constants from ALEPH hadronic tau decay data}

\author{Diogo Boito\\
Instituto de F\'isica de S\~ao Carlos, Universidade de S\~ao Paulo, CP 369,
13560-970, S\~ao Carlos, SP, Brazil\\
        E-mail: \email{boito@ifsc.usp.br}}

\author{Anthony Francis\\
        Dept. of Physics and Astronomy, York Univ., Toronto, ON Canada M3J~1P3 \\
        E-mail: \email{afrancis.heplat@googlemail.com}}
        
\author{\speaker{Maarten Golterman}\\
        Dept. of Physics and Astronomy, San Francisco State Univ.,
        San Francisco, CA 94132, USA\\
        E-mail: \email{maarten@sfsu.edu}}

\author{Renwick Hudspith\\
        Dept. of Physics and Astronomy, York Univ., Toronto, ON Canada M3J~1P3\\
        E-mail: \email{renwick.james.hudspith@gmail.com}}
        
\author{Randy Lewis\\
        Dept. of Physics and Astronomy, York Univ., Toronto, ON Canada M3J~1P3\\
        E-mail: \email{randy.lewis@yorku.ca}}
        
\author{Kim Maltman\\
        Dept. of Mathematics and Statistics, York Univ., Toronto, ON Canada M3J~1P3\\
        CSSM, Univ. of Adelaide, Adelaide, SA 5005 Australia\\
        E-mail: \email{kmaltman@yorku.ca}}
        
\author{Santiago Peris\\
        Dept. of Physics, Univ. Aut\`onoma de Barcelona, E-08193 Bellaterra, 
        Barcelona, Spain\\
        E-mail: \email{peris@ifae.es}}

\abstract{We determined the NLO chiral low-energy constant $L_{10}$, and various combinations of NNLO chiral low-energy constants employing recently revised ALEPH results for the non-strange vector ($V$) and axial-vector ($A$) hadronic tau decay distributions and recently updated RBC/UKQCD lattice data for the non-strange $V-A$ two-point function. In this talk, we explain the ingredients of this determination. Our errors are at or below the level expected for contributions of yet higher order in the chiral expansion, suggesting that our results exhaust the possibilities of what can be meaningfully achieved in an NNLO analysis.}

\FullConference{The 8th International Workshop on Chiral Dynamics, CD2015 ***\\
		29 June 2015 - 03 July 2015\\
		Pisa,Italy}

\begin{document}

\section{Introduction}
In a recent paper \cite{L10ALEPH} we revisited the determination of the
next-to-leading order (NLO)
low-energy constant (LEC) $L_{10}$, and various combinations of NNLO
LECs (denoted by $C_i$) from hadronic $\tau$ decays, which we had previously \cite{bgjmp13,GMPNNLO}
extracted from OPAL data \cite{OPAL}.   The motivation for doing so is that the
ALEPH data for the hadronic spectral functions
measured from $\tau$ decays has recently been revised \cite{ALEPH13}, to correct for the inadvertent omission 
of unfolding correlations from the covariance matrix \cite{TAU10}.   The ALEPH data
are more precise, and thus lead to values for these LECs with smaller errors.

These data were analyzed previously in Ref.~\cite{alphas14}, the main goal being an
extraction of $\alpha_s(m_\tau^2)$, and we will make use of the results of that analysis here.
These data are used in finite-energy sum rules (FESRs) involving the non-strange $V-A$ combinations
of vector two-point correlators and spectral functions. We also employ flavor-breaking inverse moment
FESRs in the $V$ and $V\pm A$ channels, following the pioneering work of Ref.~\cite{DK}, since
these provide access to additional combinations of NNLO LECs. For these we also need data from
strange hadronic tau decays. Results for the main exclusive modes were taken from BaBar and Belle,
with ALEPH data used for the remainder (details, including references, can be found in Ref.~\cite{GMPNNLO}).

Finally, we also employ lattice data \cite{L10ALEPH,Boyleetal14,rbcukqcdfine11,rbcukqcdcoarse12,rbcukqcdphyspt14}, since the lattice
allows us to vary the pseudo Nambu-Goldstone boson masses, and thus to
separate, in particular, $L_{10}$ from $C_{12}-C_{61}+C_{80}$ and
$C_{13}-C_{62}+C_{81}$.   The values of $L_5$ and $L_9$, which also contribute
to the $V$ and $A$ two-point correlators \cite{ABT}, are input to our analysis,
and taken from Refs.~\cite{MILC9A} and \cite{bt02}, respectively.

The summary of our work which we will present here will be brief.   All details
can be found in Ref.~\cite{L10ALEPH} and references therein.

\section{Sum rules}
We begin with the non-strange $V-A$ vacuum polarization sum rules.
Define
\begin{equation}
\label{VmAvacpol}
\overline\Pi_{V-A}^{(w)}(Q^2)=\int_0^\infty ds\,w(s/s_0)\,\frac{\rho_V(s)-\overline\rho_A(s)}{s+Q^2}\ ,\qquad 0<s_0\le m_\tau^2\ ,
\end{equation}
with $\rho_{V/A}$ the $V/A$ non-strange spectral function, with a bar indicating that
the contribution from the pion has been omitted, and where $w(x)$ is a polynomial.
Then \cite{bgjmp13}
\begin{eqnarray}
\label{Leff}
-8L_{10}^{\rm eff}&\equiv&\overline\Pi_{V-A}(0)\\&=&
\overline\Pi_{V-A}^{(w_2)}(0)+\frac{4f_\pi^2}{s_0}\Bigl(1-\underbrace{\frac{17\alpha^2_s(s_0)m^2_{u,d}(s_0)}{16\pi^4 f_\pi^2}-\frac{m_\pi^2}{2s_0}\left(1+O(\alpha_s)\right)}_{\rm numerically\ negligible}\Bigr)\ ,\nonumber\\
-16C_{87}^{\rm eff}&=&\overline\Pi'_{V-A}(0)\ ,\nonumber
\end{eqnarray}
with the choice of weight
\begin{equation}
\label{weight}
w_2(x)=(1-x)^2
\end{equation}
for the polynomial $w(x)$ in Eq.~(\ref{VmAvacpol}),
defines effective LECs $L_{10}^{\rm eff}$ and $C_{87}^{\rm eff}$, related to, but
not equal to, $L_{10}$ and $C_{87}$ (see below).   The expression with the weight $w_2$ leads to smaller error on $L_{10}^{\rm eff}$, because the second term is
known very precisely (even if we neglect the terms labeled as ``numerically
negligible'').   We will therefore employ this expression to obtain the numerical value of
$L_{10}^{\rm eff}$.

We extract $\overline\Pi_{V-A}^{(w)}(Q^2)$ from the data using the split
\begin{eqnarray}
\label{vacpoldata}
\overline\Pi^{(w)}_{V-A}(Q^2)&=&\sum_{{\rm bins}<s_{\rm sw}}
\!\!\!\!\!w_{\rm av}\left(\frac{s_{\rm bin}}{s_{\rm sw}}\right)
\frac{\rho_V(s_{\rm bin})-\overline\rho_A(s_{\rm bin})}{s_{\rm bin}+Q^2}\\
&&+\int_{s_{\rm sw}}^\infty ds\,w\left(\frac{s}{s_{\rm sw}}\right)\frac{\rho_V^{\rm DV}(s)-\rho_A^{\rm DV}(s)}{s+Q^2}\ ,\nonumber
\end{eqnarray}
where the first term is determined directly from the data,\footnote{The subscript
``av'' on $w$ indicated that we average the weight $w$ over the width of each bin.}
and the second term is evaluated using the parametrization
\begin{equation}
\label{DVansatz}
\rho_{T}(s)=e^{-\delta_T-\gamma_T s}\sin(\alpha_T+\beta_T s)\ ,\qquad T\in\{V,\ A\}\ .
\end{equation}
Values for the parameters in Eq.~(\ref{DVansatz}) were obtained from fits to
weighted moments of the ALEPH spectral functions in Ref.~\cite{alphas14}.
Furthermore, we took $s_{\rm sw}=1.55$~GeV$^2$, and all correlations, including those
between the first and second terms in Eq.~(\ref{vacpoldata}) were taken into
account.\footnote{At $Q^2=0$ there is no discernible sensitivity to the precise
value of $s_{\rm sw}$.}   While the contribution of the second term in Eq.~(\ref{vacpoldata}) is small, including it allows us to check on the contribution
from duality violations to
$L_{10}^{\rm eff}$ and $C_{87}^{\rm eff}$ quantitatively. 

For the flavor-breaking inverse-moment sum rules, we define
\begin{equation}
\label{DeltaPi}
\Delta\Pi_T(Q^2)\equiv\Pi_{ud;T}(Q^2)-\Pi_{us;T}(Q^2)\ ,\qquad T\in\{V,\ A,\ V\pm A\}\ ,
\end{equation}
and the sum rules of interest then take the form \cite{L10ALEPH,DK}
\begin{eqnarray}
\label{FBIMFESR}
\Delta\Pi_V(0)&=&\int_{4m_\pi^2}^{s_0}\!\!\!\!\!ds\,\frac{w(s/s_0)}{s}\,\Delta\rho_V(s)+\mbox{OPE}\ ,\\
\Delta\overline\Pi_{V\pm A}(0)&=&\int_{4m_\pi^2}^{s_0}\!\!\!\!\!ds\,\frac{w(s/s_0)}{s}\,\Delta\overline\rho_{V\pm A}(s)
\pm \left(\frac{2f_K^2}{m_K^2}(1-w\left(\frac{m_K^2}{s_0}\right))-(K\to\pi)\right)+\mbox{OPE}\ ,\nonumber\\
w(x)&=&\begin{cases}(1-x)^3 \\ (1-x)^3(1+x+\frac{1}{2}x^2)
\qquad(\mbox{DK\ weight}) 
\end{cases}\ .\nonumber
\end{eqnarray}
In this case, we neglected the contribution from duality violations, because of the fact
that both weights are triply pinched at $s=s_0$ and the additional $1/s$ suppression
of the region near $s=s_0$.   We included perturbative and non-perturbative contributions from the operator product expansion
(OPE) with conservative estimates of the systematic errors for these contributions.  As clearly the left-hand side of these equations
is independent of $s_0$, this allowed us to carry out a self-consistency check by considering the
$s_0$ dependence of the right-hand side, which we did on the interval
$2$~GeV$^2<s_0<m_\tau^2$.  We found that $s_0$-independence is well satisfied
within errors.
In addition, we checked that our results
are independent which of the two weights in Eq.~(\ref{FBIMFESR}) was
employed.

We summarize the results we find from employing these sum rules:
\begin{eqnarray}
\label{resultsQ20}
L_{10}^{\rm eff}&=&-6.446(50)\times 10^{-3}\ ,\\
C_{87}^{\rm eff}&=&8.38(18)\times 10^{-3}\ \mbox{GeV}^{-2}\ ,\nonumber\\
\Delta\Pi_V(0)&=&0.0224(9)\ ,\nonumber\\
\Delta\overline\Pi_A(0)&=&0.113(8)\ ,\nonumber\\
\Delta\overline\Pi_{V+A}(0)&=&0.0338(10)\ ,\nonumber\\
\Delta\overline\Pi_{V-A}(0)&=&0.0111(11)\ .\nonumber
\end{eqnarray}

\section{Connection with chiral perturbation theory}
The connection between the effective LECs of Eq.~(\ref{Leff}) and chiral
perturbation theory (ChPT) is given by \cite{ABT}
\begin{eqnarray}
L_{10}^{\rm eff}&=&L_{10}^r\left(1-4(2\mu_\pi+\mu_K)\right)
-2(2\mu_\pi+\mu_K)L_9^r-\frac{1}{8}\hat{R}_{\pi K}(\mu,0)\label{connectionL10}\\
&&-4m_\pi^2(C_{12}^r-C_{61}^r+C_{80}^r)-4(2m_K^2+m_\pi^2)(C_{13}^r-C_{62}^r+C_{81}^r)\ ,\nonumber\\
C_{87}^{\rm eff}&=&C_{87}^r-\frac{1}{64\pi^2 f_\pi^2}\left(1-\log\frac{\mu^2}{m_\pi^2}+\frac{1}{3}\log\frac{m_K^2}{m_\pi^2}\right)L_9^r-\frac{1}{16}\hat{R}'_{\pi K}(\mu,0)\ ,
\label{connectionC87}
\end{eqnarray}
where $\mu_P=(m_P^2/(32\pi^2 f_\pi^2))\log(m_P^2/\mu^2)$, and 
the functions $\hat{R}_{\pi K}$ and $\hat{R}'_{\pi K}$ are known functions of the renormalization
scale $\mu$ \cite{L10ALEPH}, $m_\pi$, and $m_K$.   Below, we will take $\mu=770$~MeV, and
$L_9^r=5.93(43)\times 10^{-3}$ \cite{bt02}.

Clearly, the only way to disentangle $L_{10}^r$ from the combinations $C_{12}^r-C_{61}^r+C_{80}^r$
and $C_{13}^r-C_{62}^r+C_{81}^r$ is to vary the pion (or the kaon) mass, and
this can only be done using Lattice QCD.   To this end, we employ three 
RBC/UKQCD ensembles, two with $1/a=1.379(7)$~GeV and $m_\pi=172$ or
$250$~MeV, and one with $1/a=1.785(5)$~GeV and $m_\pi=340$~MeV 
\cite{L10ALEPH,Boyleetal14,rbcukqcdfine11,rbcukqcdcoarse12,rbcukqcdphyspt14}.

For the $\Delta\Pi_V(0)$ and $\Delta\overline\Pi_{V\pm A}(0)$, the ChPT expressions, substituting physical masses and decay constants, 
setting $\mu=770$~MeV, and defining $\Delta_{\pi K}\equiv 32(m_K^2-m_\pi^2)
=7.238$~GeV$^2$,
are given by
\begin{eqnarray}
\label{DeltaPiChPT}
\Delta\Pi_V(0)&=&0.00775-0.7218L_5^r+1.423L_9^r+1.062L_{10}^r+\Delta_{\pi K}C_{61}^r\ ,\\
\Delta\overline\Pi_{V+A}(0)&=&0.00880-0.7218L_5^r+1.423L_9^r+\Delta_{\pi K}(C_{12}^r+C_{61}^r+C_{80}^r)\ ,\nonumber\\
\Delta\overline\Pi_{V-A}(0)&=&0.00670-0.7218L_5^r+1.423L_9^r+2.125L_{10}^r
-\Delta_{\pi K}(C_{12}^r-C_{61}^r+C_{80}^r)\ .\nonumber
\end{eqnarray}
Using the value $L_5^r=0.84(38)\times 10^{-3}$ from Ref.~\cite{MILC9A} and 
the value for $L_9^r$ quoted above, these expressions allow us to obtain
$C_{12}^r+C_{61}^r+C_{80}^r$ from $\Delta\overline\Pi_{V+A}(0)$, then
from $\Delta\overline\Pi_{V-A}(0)$, $L_{10}^{\rm eff}$ and the lattice we get
$C_{12}^r-C_{61}^r+C_{80}^r$ and $C_{13}^r-C_{62}^r+C_{81}^r$, while
$\Delta\Pi_V(0)$ gives us direct access to $C_{61}^r$.

Taking all correlations into account, this set up yields the following results
for $L_{10}^r$ and several (combinations of) NNLO LECs\footnote{For the
many details, we refer to Refs.~\cite{L10ALEPH,bgjmp13,GMPNNLO}.}
\begin{eqnarray}
\label{endresults}
L_{10}^r&=&-3.50(17)\times 10^{-3}\ ,\\
C_{12}^r+C_{61}^r+C_{80}^r&=&\phantom{-}2.37(16)\times 10^{-3}\ \mbox{GeV}^{-2}\ ,\nonumber\\
C_{12}^r-C_{61}^r+C_{80}^r&=&-0.56(15)\times 10^{-3}\ \mbox{GeV}^{-2}\ ,\nonumber\\
C_{13}^r-C_{62}^r+C_{81}^r&=&\phantom{-}0.46(9)\times 10^{-3}\ \mbox{GeV}^{-2}\ ,\nonumber\\
C_{61}^r&=&\phantom{-}1.46(15)\times 10^{-3}\ \mbox{GeV}^{-2}\ ,\nonumber\\
C_{12}^r+C_{80}^r&=&\phantom{-}0.90(9)\times 10^{-3}\ \mbox{GeV}^{-2}\ ,\nonumber\\
C_{87}^r&=&\phantom{-}5.10(22)\times 10^{-3}\ \mbox{GeV}^{-2}\ ,\nonumber
\end{eqnarray}
where $C_{87}^r$ was obtained directly from Eq.~(\ref{connectionC87}).

\section{Conclusion}
The results presented in Eq.~(\ref{endresults}) constitute our best results for these
(combinations of) LECs.   They are consistent with the results obtained earlier
from OPAL data in Refs.~\cite{bgjmp13,GMPNNLO,Boyleetal14}, but the errors are smaller.
Since our analysis method was the same, this is due to the higher precision of
the revised ALEPH data.

Of course, one wonders whether it is possible to do better.   However, there is a
systematic effect due to the neglect of N$^3$LO terms in ChPT, which is not 
included in the errors shown in Eq.~(\ref{endresults}).   We estimate that the 
error due to neglecting higher orders in ChPT is about 6\% for $L_{10}^r$, and
about 25\% for NNLO LECs.\footnote{For a detailed discussion of this, see 
Ref.~\cite{bgjmp13}.}   This means that the precision attainable with an NNLO
analysis has been reached.

We note that the combination $C_{13}^r-C_{62}^r+C_{81}^r$ is not smaller
in size than the combination $C_{12}^r-C_{61}^r+C_{80}^r$, even though the
LECs $C_{13,62,81}^r$ are suppressed relative to $C_{12,61,80}^r$ in the
limit $N_c\to\infty$.   Since cancellations in the combination $C_{12}^r-C_{61}^r+C_{80}^r$ take place (note that the values for $C_{12}^r+C_{80}^r$
and $C_{61}^r$ are indeed significantly larger than that for $C_{12}^r-C_{61}^r+C_{80}^r$),
this does not invalidate the expansion in $1/N_c$; however, it does invalidate
the assumption made in Ref.~\cite{GPP} that the combination $C_{13}^r-C_{62}^r+C_{81}^r$ can be set approximately equal to zero.   We also note that our parametrization
of the duality-violating part of the spectral functions, given in Eq.~(\ref{DVansatz}),
is more general than that employed in Ref.~\cite{GPP} (whereas in Ref.~\cite{Dominguezetal} they were neglected altogether), thus
avoiding unnecessary additional assumptions.   For a detailed discussion of the
assumptions underlying the use of Eq.~(\ref{DVansatz}), we refer to Ref.~\cite{US2}.

Finally, we remark that in Ref.~\cite{L10ALEPH} we also determined the 
coefficients of $Q^{-6}$ and $Q^{-8}$ in the OPE of the $V-A$ two-point function.   We find values that differ
by about $2.4$ standard deviations from those we obtained from the OPAL data in
Ref.~\cite{bgjmp13}.\\

\noindent {\bf Acknowledgments}
\vspace{3ex}

DB thanks the  Department of Physics of the
Universitat Aut\`onoma de Barcelona, and 
KM and SP thank the Department of Physics and Astronomy at San 
Francisco State University for hospitality. 
MG is supported in part by the US Department of Energy, AF, RH, RL and KM are
supported by grants from the Natural Sciences and Engineering Research 
Council of Canada, and SP is supported by CICYT-FEDER-FPA2014-55613-P, 
2014~SGR~1450, the Spanish Consolider-Ingenio 2010 Program
CPAN (CSD2007-00042). Propagator inversions for the improved
lattice data were performed on the STFC funded ``DiRAC'' BG/Q
system in the Advanced Computing Facility at the University of
Edinburgh.

\end{document}